\documentclass[english,reprint,sort&compress]{revtex4-1}
\usepackage[T1]{fontenc}
\usepackage[utf8]{inputenc}
\usepackage{textcomp}
\usepackage{amsmath}
\usepackage{amssymb}
\usepackage{graphicx}
\usepackage{esint}
\usepackage{subscript}

\makeatletter
\usepackage{hyperref}
\usepackage{amssymb,amsmath}

\makeatother

\usepackage{babel}
\begin{document}
\title{Sequential lithium deposition on hexagonal boron nitride monolayer
on Ir(111): Identifying intercalation and adsorption}
\author{Marin Petrovi\'{c}}
\email{mpetrovic@ifs.hr}

\affiliation{Center of Excellence for Advanced Materials and Sensing Devices, Institute
of Physics, Bijenička 46, 10000 Zagreb, Croatia}
\begin{abstract}
Stepwise deposition of Li atoms onto hexagonal boron nitride (hBN)
monolayer on Ir(111) is investigated by means of angle-resolved photoemission
spectroscopy and low-energy electron diffraction. Sequential Li deposition
progressively shifts the band structure of hBN to higher binding energies
due to the induction of a variable electric potential originating
from electronic charge donation by alkali atoms. At small coverages,
Li atoms preferably intercalate under the hBN layer, where they are
highly charged and give rise to a large initial shift of the band
structure. Additionally, intercalated Li atoms effectively decouple
hBN from the substrate and consequently reduce its moiré corrugation.
As the deposition progresses further, Li atoms also adsorb on top
of hBN, and the average effective charge of both intercalated and
adsorbed Li atoms progressively diminishes due to the Coulomb repulsion
penalty, which is partially screened by the metal substrate in the
intercalated subsystem. This gives rise to a saturation of the respective
electric potential and the band structure shift, and elaborates on
the pathways and limitations of chemical functionalization of epitaxial
hBN systems with charge donating species.
\end{abstract}
\keywords{Keywords: hexagonal boron nitride, alkali metals, electrostatic gating,
ARPES, LEED}
\maketitle

\section{Introduction}

There are many intrinsic properties of hexagonal boron nitride (hBN),
such as large band gap, chemical inertness and atomic-level flatness
\citep{Watanabe2004,Corso2004,Pakdel2014}, that make it very interesting
subject of scientific research and a promising material for future
applications. The investigation of hBN gained additional momentum
after the realization of layered van der Waals heterostructures and
the respective devices, such as transistors \citep{Dean2010a,Lee2015},
light-emitting diodes \citep{Ross2014}, gas sensors \citep{Sajjad2013}
and solar cells \citep{Lin2015p}, where the properties of hBN turned
out to be crucial for those complex systems as a whole. In stacked
heterostructures, hBN is in close contact with other substances and
is often subjected to electric or magnetic fields, and in order to
fully understand and exploit those systems, a thorough investigation
of all relevant interactions is needed.

Interlayer interactions are certainly crucial and need to be addressed
on a fundamental level. This can be achieved by utilizing a surface
science approach, i.e., by fabricating well-defined (mono)layers of
hBN in ultra-high vacuum (UHV), followed by their controlled decoration
with atoms or molecules of interest \citep{Auwarter2019}. Hereafter,
we focus on materials which act as charge donors or acceptors, and
as such are able to significantly alter the electronic properties
of hBN. The simplest, yet very efficient charge donors are the alkali
atoms. Their effect on the electronic structure of hBN has been investigated
in several studies, where one of the most notable observations is
the shift of the electronic bands to higher binding energies. Such
shift arises from the electric potential which is induced by the charge
transferred form alkali atoms to their surroundings. Fedorov \textit{et
al}. found that K and Li deposition on hBN/Au/Ni(111) results in two
different structures: Li remains adsorbed on top of hBN and causes
a shift of the valence bands of 0.9 eV, while K intercalates under
the hBN and induces a shift of 2.77 eV \citep{Fedorov2015}. Cai \textit{et
al}. conducted Cs deposition on hBN/Ir(111), and identified two Cs
configurations, adsorbed and a combination of adsorbed and intercalated,
inducing valence band shifts of 0.35 and 3.25 eV, respectively \citep{Cai2018}.
Besides the valence bands, B and N core levels have also shifted to
higher binding energies in these two studies. By investigating a somewhat
different system comprised of multilayer hBN on TiO\textsubscript{2}(100),
Koch \textit{et al}. measured a 2.5 eV shift of the valence bands
to higher binding energies after K deposition \citep{Koch2018}. In
an analogous way to alkali metals, deposition of charge acceptor species
on epitaxial hBN causes shift of the electronic bands to lower binding
energies, as has been shown for molecular oxygen adsorption on hBN/Ni(111),
where the valence bands shifted by 1.2 eV closer to the Fermi level
\citep{Spath2019}.

Being the smallest of alkali metals, Li is an interesting candidate
for adsorbtion on and intercalation of hBN mono- or multi-layers at
a wide range of concentrations, which could enable realization of
interesting hBN-based systems. For example, it has been demonstrated
that Li-functionalized hBN has a potential to serve as an electrode
in batteries \citep{Zhang2016k}. Also, calculations predict that
Li atoms encapsulated by two hBN layers could be suitable for hosting
plasmonic excitations \citep{Loncaric2018}. Li adsorbed on monolayer
hBN could potentially invoke n-type conductivity and expedite integration
of hBN into electronics \citep{Ding2016}.

In this work, we further elaborate on the effects of hBN decoration
with Li atoms. More specifically, angle-resolved photoemission spectroscopy
(ARPES) and low-energy electron diffraction (LEED) are utilized to
investigate the electronic and morphological characteristics of hBN
on Ir(111) at different Li concentrations and with that, at variable
electronic charge arrangement. Sequential Li deposition employed in
our experiments reveals the pathway for Li intercalation and adsorption,
and allows for a detailed investigation of charge transfer dynamics.

\section{Methods}

Sample preparation and all experimental measurement were conducted
in an ultra-high vacuum setup (base pressure of $\approx5\times10^{-10}$
mbar) dedicated to ARPES, with the LEED instrument available as an
auxiliary technique. Ir(111) single crystal cleaning consisted of
repeated cycles of Ar ion sputtering at room temperature (RT) at 1.5
keV energy, oxygen glowing ($p=10^{-6}$ mbar) at 1170 K and annealing
at 1470 K. The growth of hBN proceeded by exposing Ir(111) to borazine
(B\textsubscript{3}H\textsubscript{6}N\textsubscript{3}, $p=2\text{\texttimes}10^{-7}$
mbar) at 1170 K for 15 minutes. Keeping the sample temperature below
1220 K at all times prevented decomposition of hBN and appearance
of epitaxial boron \citep{Petrovic2017a}. The quality of hBN was
checked by ARPES and LEED, where well-defined $\pi$ and $\sigma$
bands, and the pronounced moiré diffraction spots have been sought.
Li was deposited in a sequence of steps from commercial dispensers
(SAES getters) at RT. Prior to each Li deposition sequence, a fresh
hBN sample has been synthesized.

ARPES measurements were carried out at RT with a Scienta SES 100 analyzer
(25 meV energy resolution, 0.2$^{\circ}$ angular resolution). Data
has been collected in the $\mathrm{\Gamma K}$ direction. A helium
discharge lamp ($h\nu=21.2$ eV, non-polarized) was utilized as a
photon source, with the spot diameter on the sample of $\approx2$
mm.

\section{Results}

\begin{figure}
\begin{centering}
\includegraphics{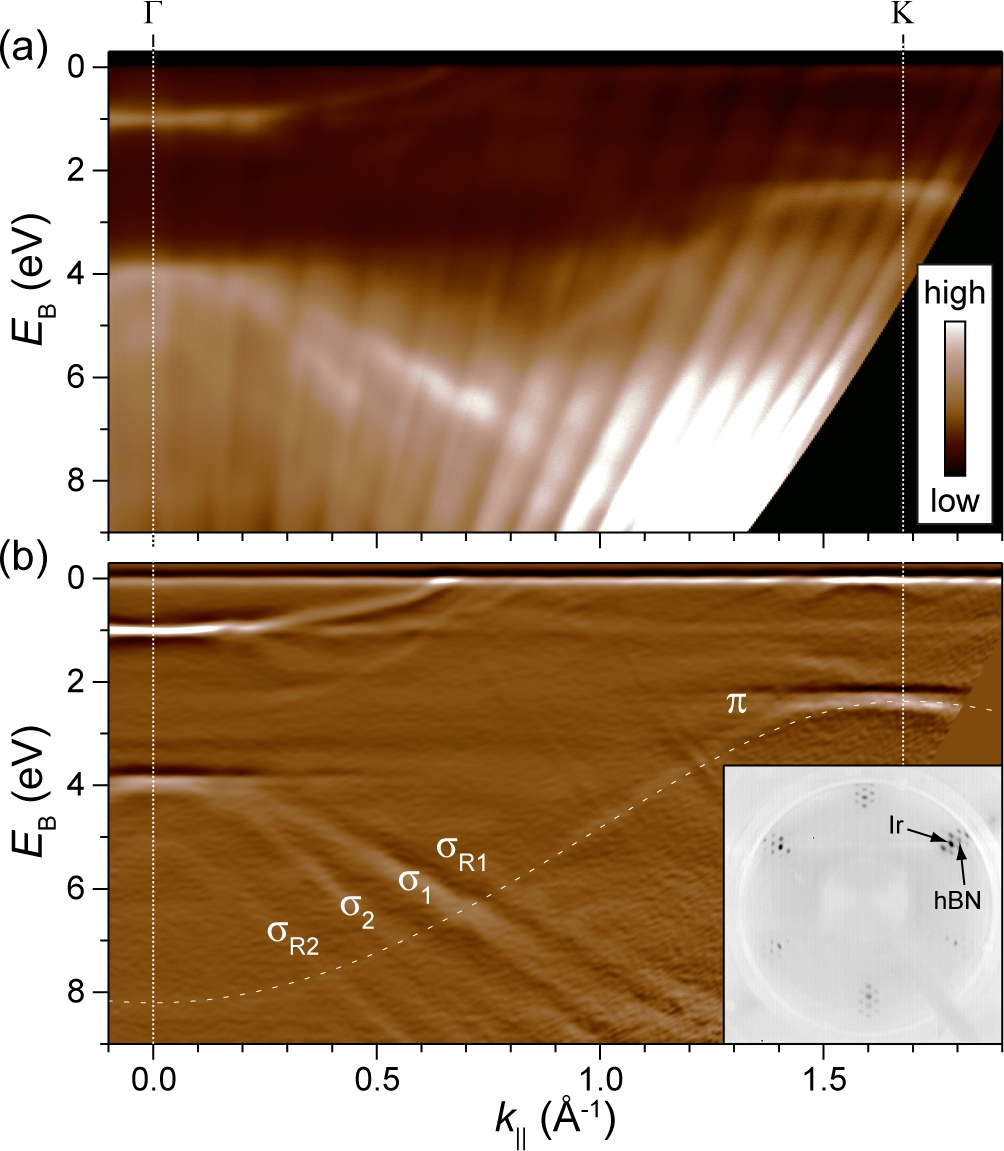}
\par\end{centering}
\caption{\label{fig1} ARPES map of hBN/Ir(111) system along the $\Gamma\mathrm{K}$
direction presented as (a) raw data and (b) second derivative in the
$y$ coordinate. Electronic bands of hBN, $\pi$ and the two $\sigma$
bands ($\sigma_{1}$ and $\sigma_{2}$) are visible. Additionally,
replicated $\sigma$ bands ($\sigma_{\mathrm{R1}}$ and $\sigma_{\mathrm{R2}}$)
can be discerned as well. Thin dashed line in panel (b) is the TBA
fit to the $\pi$ band. The inset in panel (b) shows LEED image of
the system ($E=56$ eV), with the moiré diffraction spots surrounding
the first order diffraction spots of Ir and hBN.}
\end{figure}

As a starting point for Li deposition experiments, epitaxial hBN samples
were prepared. ARPES mapping of hBN/Ir(111) in the $\Gamma\mathrm{K}$
direction, shown in Fig. \ref{fig1}, provides proof of good quality
of hBN, with $\sigma_{1}$, $\sigma_{2}$ and $\pi$ bands visible.
The replicated $\sigma$ bands ($\sigma_{\mathrm{R1}}$ and $\sigma_{\mathrm{R2}}$)
are also evident, indicating that a well-defined moiré corrugation
is present in the hBN/Ir(111) system \citep{Usachov2012}, again signaling
for a very good uniformity of hBN over mesoscopic scales. For the
sake of completeness, we fit the measured $\pi$ band with the first-nearest-neighbor
tight binding approximation (1NN TBA) \citep{Sawinska2010,Ribeiro2011a}.
Such fitting which provides boron and nitrogen onsite energies of
3.73 and -2.37 eV, respectively, and the hopping energy between the
nearest neighbors of 2.78 eV (see Appendix A for more details). The
fitted $E_{\mathrm{TBA}}\left(k_{\parallel}\right)$ dispersion is
plotted in Fig. \ref{fig1}(b) by a dashed line. Crystallographic
quality of hBN has been confirmed by LEED data {[}see inset in Fig.
\ref{fig1}(b){]}, where the diffraction satellite spots of the moiré
structure surround the first order Ir and hBN spots.

\begin{figure*}
\begin{centering}
\includegraphics{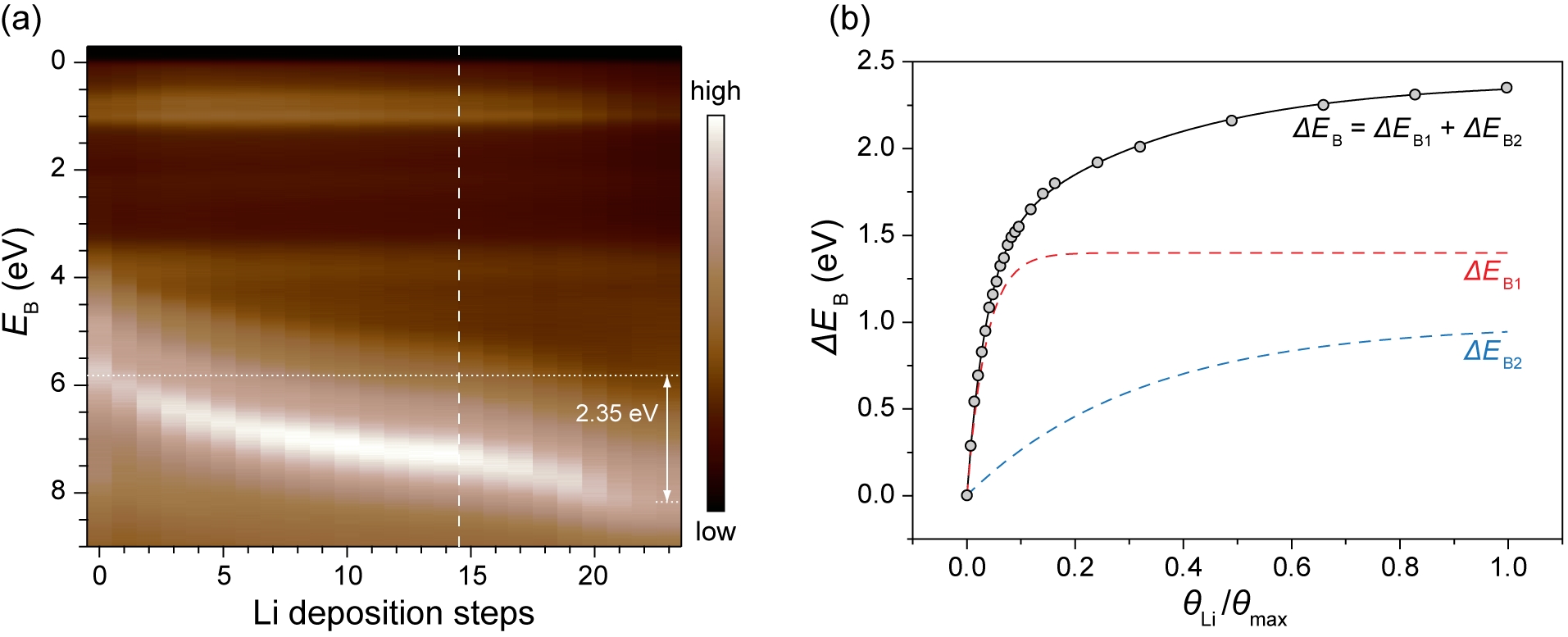}
\par\end{centering}
\caption{\label{fig2}(a) A stack of EDCs at $k_{\parallel}=0.4$ Å\protect\textsuperscript{-1}
for a sequence of Li deposition steps. 1-minute-long (left of dashed
line) and 2-, 4- and 6-minutes-long deposition steps (right of dashed
line) have been employed. The shifting bands are the $\sigma$ bands
of hBN. Non-shifting Ir bands are also visible. (b) Increase of the
$\sigma$ bands binding energy, $\Delta E_{\mathrm{B}}$, as a function
of Li coverage, $\theta_{\mathrm{Li}}$ (gray circles). The data has
been fitted (full black curve) by a sum of the two exponential functions
$\Delta E_{\mathrm{B1}}$ and $\Delta E_{\mathrm{B2}}$ (dashed red
and blue curves).}
\end{figure*}

Li deposition on hBN/Ir(111) has been conducted in a series of steps.
Due to the limited photon energy and the respective restrictions in
the size of $k$-space available in our experiments, focus is put
on the $\sigma$ bands and evolution of their binding energy as a
function of the deposited Li amount ($\theta_{\mathrm{Li}}$). A stack
of energy distribution curves (EDCs) extracted at $k_{\parallel}=0.4$
Å\textsuperscript{-1} is shown in a color plot in Fig. \ref{fig2}(a).
At first, Li was deposited in 1-minute-long steps, and subsequently
in 2-, 4- and 6-minutes-long steps {[}data right of the dashed line
in Fig. \ref{fig2}(a){]} in order to reach the maximum shift of the
$\sigma$ bands more efficiently. The downshift of the $\sigma$ bands,
i.e., the increase of their binding energies is evident. Initially,
the shift progresses relatively fast, but afterwards it slows down
and eventually saturates at a value of 2.35 eV. This is visible from
the comparison of $\sigma$ band position in consecutive EDCs: employment
of 1-minute-long steps induces progressively diminishing shift (up
to the deposition step 14), and only utilization of longer, i.e.,
2-, 4- and 6-minutes-long steps (deposition step 15 and beyond) allows
for a distinction of further shift of the bands. Such observation
points to a non-linear shift of the bands as a function of number
of Li deposition steps and the cumulative amount of deposited Li.
Application of additional Li deposition steps, beyond the ones shown
in \ref{fig2}(a), did not produce further shift of the $\sigma$
bands. Instead, it resulted in blurring and reduction of photoemission
intensity of the $\sigma$ bands, which might stem from Li multilayer
formation on top of hBN.

Electronic bands that do not show a change in binding energy in Fig.
\ref{fig2}(a) originate from Ir. The most notable bands are the ones
located at $\approx1$ eV below the Fermi level. These bands show
a decrease in intensity as the number of Li deposition steps increases,
which can be explained by attenuation of Ir photoelectrons due to
passage through additional Li layers (similar trend is found for the
$\sigma$ bands intensity). It should be noted that these states are
Ir bulk states and not surface states \citep{Pletikosic2010}, and
therefore their attenuation cannot be related to interaction between
Ir surface atoms and deposited Li atoms.

The binding energy increase of the $\sigma$ bands, $\Delta E_{\mathrm{B}}$,
has been determined for each deposition step by fitting the corresponding
EDC curve with a Lorentzian lineshape. Furthermore, the amount of
deposited Li on the sample after each step has been calculated (i)
by adopting a linear relation between Li dispenser yield and the deposition
time, and (ii) by defining that $\theta_{\mathrm{Li}}=\theta_{\mathrm{max}}$
induces a maximum shift of the $\sigma$ bands observed in our experiments
(see Appendix B for more details). With this data, it is possible
to plot $\Delta E_{\mathrm{B}}$ as a function of $\theta_{\mathrm{Li}}$,
as shown in Fig. \ref{fig2}(b) by gray circles, and such calibration
can be used for a straightforward estimation of $\theta_{\mathrm{Li}}$
in other experiments based on the binding energy of hBN bands.

Due to technical restrictions, work function measurements could not
be performed in our setup. It is therefore not possible to disclose
whether $\Delta E_{\mathrm{B}}$ arises merely from the Li-induced
modification of the sample work function and alignment of hBN bands
to the vacuum level. Hence, we consider the local electric potential
induced by Li dipoles that acts on hBN ($\phi_{\mathrm{loc}}$) for
a proper description of our system. This potential is directly proportional
to the band shift, $\Delta E_{\mathrm{B}}\sim\phi_{\mathrm{loc}}$
\citep{Fedorov2015,Cai2018}, implying that Fig. \ref{fig2}(b) can
also be interpreted as $\phi_{\mathrm{loc}}\left(\theta_{\mathrm{Li}}\right)$
graph.

Saturation of data in Fig. \ref{fig2}(b) suggests exponential-like
shift of the hBN bands as Li deposition progresses. Fitting with a
single exponential function yields poor results, but a fit with a
two-component exponential function of the form $\Delta E_{\mathrm{B1}}+\Delta E_{\mathrm{B2}}=A_{1}\left(1-e^{-\theta_{\mathrm{Li}}/\theta_{1}}\right)+A_{\mathrm{2}}\left(1-e^{-\theta_{\mathrm{Li}}/\theta_{2}}\right)$
provides an excellent agreement with the data, as indicated by a full
line in Fig. \ref{fig2}(b). Dashed lines in Fig. \ref{fig2}(b) denote
the two components of the fitting function, $\Delta E_{\mathrm{B1}}$
and $\Delta E_{\mathrm{B2}}$, where the first one exhibits rapid
growth ($\theta_{1}=0.04\:\theta_{\mathrm{max}}$) and saturation
at $A_{1}=1.40$ eV, while the second component rises more slowly
($\theta_{2}=0.32\:\theta_{\mathrm{max}}$) and saturates at $A_{2}=0.99$
eV. At low $\theta_{\mathrm{Li}}$, the $\Delta E_{\mathrm{B1}}$
component provides dominant contribution to $\Delta E_{\mathrm{B}}$,
and after $\theta_{\mathrm{Li}}\approx0.2\:\theta_{\mathrm{max}}$,
the $\Delta E_{\mathrm{B2}}$ component becomes the source of further
shift of the $\sigma$ bands.

\begin{figure*}
\begin{centering}
\includegraphics{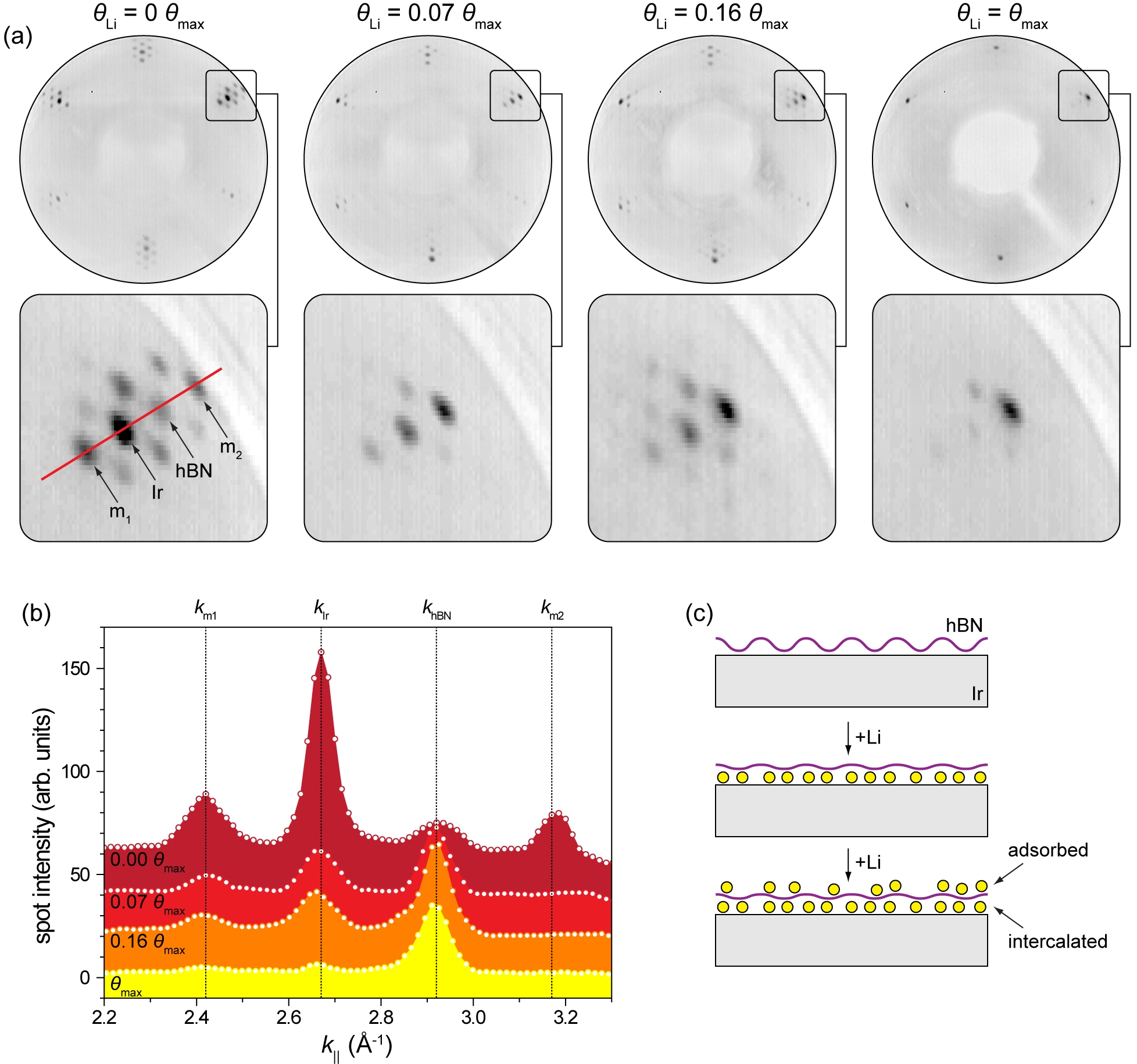}
\par\end{centering}
\caption{\label{fig3}(a) A sequence of LEED images (top panel) along with
their zoom-ins (bottom panels) at several characteristic Li coverages,
$\theta_{\mathrm{Li}}$, indicated above. Diffraction spots of Ir,
hBN, and the moiré structure (m\protect\textsubscript{1} and m\protect\textsubscript{2})
are noted. $E=56$ eV. (b) Cross-section through LEED images shown
in (a) as indicated by a red line. The curves have been shifted in
the $y$ direction for clarity. (c) A schematic model of hBN on Ir(111)
without Li (top), at intermediate Li coverage (middle), and at a maximum
Li coverage $\theta_{\mathrm{max}}$ (bottom).}
\end{figure*}

During sequential Li deposition, the diffraction pattern of the sample
has been inspected several times, as shown in Fig. \ref{fig3}(a),
upper panels. At $\theta_{\mathrm{Li}}=0\:\theta_{\mathrm{max}}$,
a diffraction pattern corresponding to the epitaxially aligned hBN
on Ir(111) is observed, along with the moiré diffraction spots surrounding
the first order hBN and Ir spots. A more detailed view of the moiré
spots is given in the lower, zoom-in panels in Fig. \ref{fig3}(a).
Good visibility of the moiré spots indicates significant corrugation
of the hBN layer \citep{Usachov2012}. A cross-section through the
first order hBN and Ir spots, as well as through the two closest moiré
spots is given in Fig. \ref{fig3}(b). After deposition of only 0.07
$\theta_{\mathrm{max}}$ of Li, intensity of the diffraction spots
changes notably: Ir and moiré spots are significantly reduced, and
the intensity of the hBN spot has increased. Additional Li deposition
leads to further reduction of Ir and moiré spots, while hBN spot intensity
remains approximately constant. At $\theta_{\mathrm{Li}}=\theta_{\mathrm{max}}$,
moiré and Ir spots are barely visible, as evident in Figs. \ref{fig3}(a)
and (b).

It should be noted that no Li superstructures were observed at any
Li coverage. Also, an increase of the Ir spot intensity, which might
indicate the formation of an intercalated $\mathrm{Li-}\left(1\times1\right)$
superstructure \citep{Pervan2015,Silva2019}, has not been registered.
Hence, we conclude that Li most likely forms disordered structures
on hBN/Ir(111) at room temperature, which is in line with Li intercalation
in bulk hBN \citep{Sumiyoshi2012}.

\begin{figure}
\begin{centering}
\includegraphics{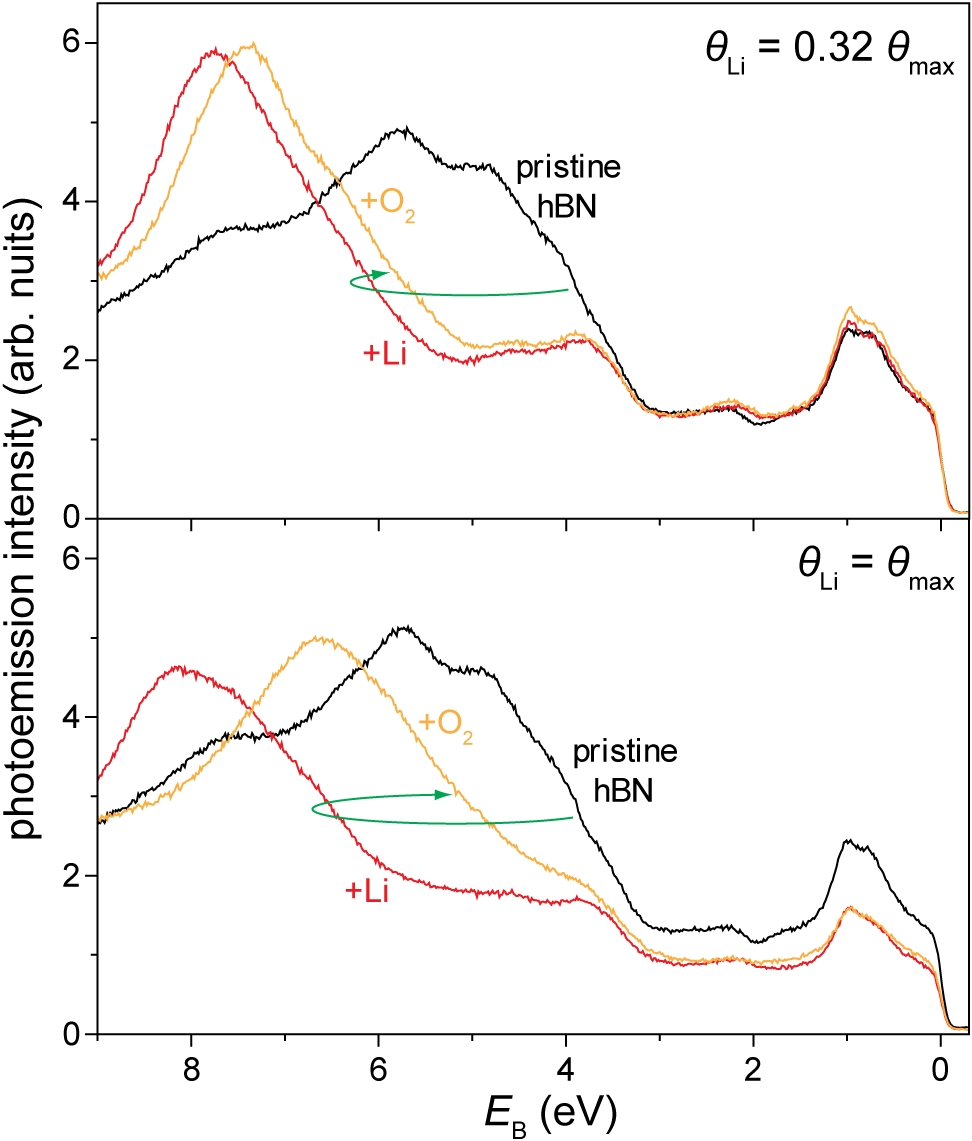}
\par\end{centering}
\caption{\label{fig4}EDCs at $k_{\parallel}=0.4$ Å\protect\textsuperscript{-1}
for samples with $\theta_{\mathrm{Li}}=0.32\:\theta_{\mathrm{max}}$
(top panel) and $\theta_{\mathrm{Li}}=\theta_{\mathrm{max}}$ (bottom
panel) that were subsequently exposed to molecular oxygen. Green arrows
indicate the shift of the $\sigma$ bands to higher binding energies
due to Li deposition (+Li, red curves), followed by a backshift to
the Fermi level after oxygen exposure (+O\protect\textsubscript{2},
orange curves).}
\end{figure}

To further investigate the arrangement of Li on hBN/Ir(111), $\theta_{\mathrm{Li}}=0.32\:\theta_{\mathrm{max}}$
and $\theta_{\mathrm{Li}}=\theta_{\mathrm{max}}$ samples have been
exposed to 100 L ($p=5\times10^{-7}$ mbar for 270 s) of molecular
oxygen (99.999 \% purity) at RT. Since alkali metals readily oxidize
in such environment \citep{Matyba2015}, this experiment can reveal
whether the alkali atoms are intercalated (therefore, protected from
oxygen by hBN), adsorbed on top of hBN (i.e., exposed to oxygen),
or take on some mixed intermediate configuration. The effects of O\textsubscript{2}
exposure are shown in Fig. \ref{fig4} where the corresponding EDCs
extracted at $k_{\parallel}=0.4$ Å\textsuperscript{-1} are shown.
For the $\theta_{\mathrm{Li}}=0.32\:\theta_{\mathrm{max}}$ sample,
deposited Li increased the binding energy of the $\sigma$ bands by
2 eV. Subsequent O\textsubscript{2} exposure resulted in a 0.28 eV
backshift to the Fermi level. For the $\theta_{\mathrm{Li}}=\theta_{\mathrm{max}}$
sample, deposited Li increased the binding energy of the $\sigma$
bands by 2.35 eV, i.e., by the largest amount observed in our experiments.
The following O\textsubscript{2} exposure then caused a significantly
larger backshift of 1.49 eV, as illustrated in Fig. \ref{fig4} by
green arrows. Additional 100 L O\textsubscript{2} exposure of the
$\theta_{\mathrm{Li}}=\theta_{\mathrm{max}}$ sample did not produce
any additional shift of the bands. Apparently, introduction of oxygen
triggers reduction of the local electric potential $\phi_{\mathrm{loc}}$,
and the magnitude of such reduction depends on the amount of Li that
has been deposited on the sample. In agreement with Fig. \ref{fig2}(a),
intensity of the Ir bands at $\approx1$ eV below the Fermi level
gets reduced after the deposition of $\theta_{\mathrm{max}}$ of Li.
Importantly, the intensity of these Ir bands is not restored upon
oxygen exposure.

\section{Discussion}

The outlined experiments clarify spatial arrangement of Li atoms at
different coverages. Reduction of the Ir and moiré spots and an increase
in intensity of the hBN spots, as visible in Fig. \ref{fig3}(b),
are typical signatures of decoupling related to the insertion of atoms
between 2D material and its substrate \citep{Ulstrup2014a,Pervan2015,Silva2019,Lin2018a}.
Since such LEED intensity modification has been observed already for
small amounts of Li, we conclude that Li atoms initially intercalate
between hBN and Ir, rather then stay adsorbed on the vacuum side of
hBN. 

Further elaboration of Li positioning is provided by O\textsubscript{2}
exposure experiments. Any Li atoms adsorbed on hBN are able to react
with O\textsubscript{2} molecules and form Li oxides, Li\textsubscript{2}O\textsubscript{x}
\citep{SHEK1990,Pervan2017}. In contrast to elemental Li, Li oxide
does not act as an efficient electron donor. Therefore, in a system
with Li\textsubscript{2}O\textsubscript{x} present, the total amount
of charge transferred to Ir is reduced in comparison to a system with
elemental Li. Consequently, corresponding induced dipole magnitude,
electric potential and shift of the electronic bands are also reduced.
We believe this scenario is visible in Fig. \ref{fig4}. Exposure
to O\textsubscript{2} resulted in a backshift for both $\theta_{\mathrm{Li}}=0.32\:\theta_{\mathrm{max}}$
and $\theta_{\mathrm{Li}}=\theta_{\mathrm{max}}$ samples as a result
of Li\textsubscript{2}O\textsubscript{x} formation on top of hBN.
Therefore, the observation of a backshift indicates the presence of
adsorbed Li, since hBN layer is chemically inert and its electronic
structure would not be affected by the presence of O\textsubscript{2}.
However, the backshift is significantly larger (approximately five
times) for the $\theta_{\mathrm{Li}}=\theta_{\mathrm{max}}$ sample.
This is explained by a larger quantity of Li adsorbed on hBN prior
to oxygen exposure, accompanied by a larger charge loss and a more
pronounced electric potential reduction when Li\textsubscript{2}O\textsubscript{x}
forms. Also, the fact that Ir bands remain attenuated after O\textsubscript{2}
exposure corroborate the scenario of Li oxide formation on top of
hBN, rather than oxygen-promoted removal of adsorbed Li. Intercalated
Li is protected from oxidation by hBN layer \citep{Matyba2015,Pervan2017}
and is responsible for the residual electric potential causing the
remaining shift of the $\sigma$ bands. We speculate that discrepancy
in the remaining shift for the two oxidized samples originates form
different amounts of Li\textsubscript{2}O\textsubscript{x} on each
of them, which give rise to different dielectric surroundings.

Overall, it can be concluded that Li is initially being intercalated
at the hBN-Ir interface, and starts adsorbing and accumulating on
top of hBN at the later stages of deposition {[}see Fig. \ref{fig3}(c){]}.
Similar behavior has been found for Li deposition on graphene/Ir(111)
\citep{Petrovic2013a,Pervan2015}, and is conditioned by the size
of Li atoms. Having small dimensions, Li atoms require a small amount
of energy to delaminate 2D materials from their substrate, and are
able to intercalate already at RT. In addition, Li atoms can penetrate
2D materials even through the smallest defects which might be impermeable
to other, larger atoms. Intercalation through defects is a reasonable
assumption, since the migration of a Li atom through a perfect hBN
mesh is energetically very expensive ($\approx7$ eV \citep{Oba2010}).
Intercalated Li cations are effectively screened by the Ir substrate
\citep{Petrovic2013a} (see also discussion below), which additionally
lowers their energy and promotes intercalation as a preferred system
configuration. However, as the concentration of intercalated atoms
approaches its maximum, adsorption from the vacuum side also becomes
allowed.

A stepwise deposition of alkali atoms employed in our study, which
has not been examined before for other similar systems \citep{Fedorov2015,Cai2018},
allows investigation of the respective charge transfer and hBN valence
band shift dynamics in more detail. This dynamics, in conjunction
with the conclusions about the Li positioning and the conducted fitting
of $\Delta E_{\mathrm{B}}$ with $\Delta E_{\mathrm{B1}}+\Delta E_{\mathrm{B2}}$,
is interpreted as follows. Initially, Li atoms form dilute intercalated
structures without well-defined crystallography. In such configuration,
Li atoms give a significant fraction of their 2\textit{s} electrons
to the Ir substrate, since the Li-Li distance is large and the corresponding
Coulomb repulsion penalty is negligible. Hence, at low $\theta_{\mathrm{Li}}$,
Li atoms are highly charged and they give rise to rapidly increasing
$\phi_{\mathrm{loc}}$ and $\Delta E_{\mathrm{B}}$. Accordingly,
it is plausible that the shift of the bands induced only by the intercalated
Li corresponds to the fast component $\Delta E_{\mathrm{B1}}$. As
$\theta_{\mathrm{Li}}$ increases, intercalated Li atoms become more
closely spaced. Then, it becomes energetically unfavorable for them
to be highly charged because the Coulomb repulsion between them would
be large, even though the charge of Li cations is partially screened
by the proximity of the metal substrate \citep{Petrovic2013a}. Overall,
due to the Coulomb penalty and screening, the effective charge per
intercalated Li atom reduces, which leads to the saturation of $\Delta E_{\mathrm{B1}}$.
In parallel, Li atoms become adsorbed on hBN, and also donate charge
to Ir. The mechanism of Coulomb penalty applies to them as well, only
without screening from the substrate due to the larger Li-Ir separation.
The charge that adsorbed Li atoms give to Ir induces additional electric
potential and the $\Delta E_{\mathrm{B2}}$ component of the overall
$\sigma$ band shift. According to Fig. \ref{fig2}(b), adsorbed Li
becomes the dominant source of $\Delta E_{\mathrm{B}}$ increase for
$\theta_{\mathrm{Li}}\gtrsim0.2\:\theta_{\mathrm{max}}$.

Therefore, an increase of Li coverage inevitably leads to a progressive
reduction of the charge transferred from each Li atom to Ir (in both
the intercalated and the adsorbed subsystems) and also screening of
Li atoms (in the intercalated subsystem). Such charge transfer dynamics
results in a continuously diminishing increase, and eventually saturation,
of $\phi_{\mathrm{loc}}$ and $\Delta E_{\mathrm{B}}$, as is evident
from Fig. \ref{fig2}(b). Indeed, it is expected that the Coulomb
penalty would provide exponential saturation of $\Delta E_{\mathrm{B1}}$
and $\Delta E_{\mathrm{B2}}$ (and therefore also $\Delta E_{\mathrm{B}}$),
since the charge of individual Li atoms reduces proportionally to
the total number of atoms in the system. In the presence of both intercalated
and adsorbed Li, the respective electric potentials add up and jointly
shift the electronic bands of hBN to higher binding energies, since
the corresponding electric dipole fields point in the same direction
towards Ir \citep{Fedorov2015}. Adsorbed Li atoms are not as highly
charged as intercalated atoms due to the substantial separation from
the metal substrate which hinders charge donation \citep{Fedorov2015},
and they cannot induce as large shifts of the electronic bands as
intercalated Li atoms do initially. However, the presence of adsorbed
Li atoms provides additional, slowly increasing electric potential
which pushes the $\sigma$ bands further to yield $\Delta E_{\mathrm{B}}$
of 2.35 eV.

\section{Summary}

Sequential deposition of Li on hBN/Ir(111) results in a stepwise shift
of the electronic bands of hBN to higher binding energies. The shift
is proportional to the magnitude of the electric potential acting
on the electrons of hBN, where the source of the potential are electric
dipoles arising from the charge transferred from Li to Ir. In the
initial stages of deposition, Li atoms get intercalated in a disordered
structure and decouple hBN from its substrate. Once intercalated,
Li atoms are highly charged, and they give rise to a significant (up
to 1.4 eV) and fast-progressing shift of the electronic bands. As
the deposition continues, Li also adsorbs on top of hBN, from where
it induces additional, somewhat smaller shift of the bands (up to
1 eV) that is characterized by a moderate increase rate. Overall,
the shift progresses rapidly in the beginning, slows down as the deposition
advances and eventually saturates at a maximum Li coverage studied.
The main reasons for the observed dynamics of valence band shift are
the facilitated charge transfer from intercalated Li atoms in comparison
to adsorbed Li atoms, the Coulomb repulsion penalty (for intercalated
and adsorbed Li) and screening from the substrate (for intercalated
Li). These factors all together cause progressive reduction of the
charge transferred per Li atom to Ir and reduction of the respective
electric potential. The presented results shed new light onto the
interaction of epitaxial hBN with charged species and describe the
response of its electronic bands to variable electric potentials,
and as such can be beneficial for optimization of chemical functionalization
and electric gating of hBN.
\begin{acknowledgments}
Financial support by the Center of Excellence for Advanced Materials
and Sensing Devices (ERDF Grant KK.01.1.1.01.0001) and by the Alexander
von Humboldt foundation is acknowledged.
\end{acknowledgments}

\section*{Declaration of competing interests}

The author declares no competing financial interests.

\setcounter{equation}{0}
\renewcommand{\theequation}
{A\arabic{equation}}

\section*{Appendix: TBA fitting and Li coverage calibration}

\subsection*{A. TBA fitting}

A good description of the $\pi$ bands of hBN can be obtained form
tight binding approximation (TBA). By taking into account only the
first nearest neighbors, the dispersion of the $\pi$ bands can be
described by \citep{Sawinska2010,Ribeiro2011a}

\begin{equation}
\begin{aligned}E_{\mathrm{TBA}}= & \frac{E_{\mathrm{B0}}+E_{\mathrm{N0}}}{2}\pm\frac{1}{2}\sqrt{\left(E_{\mathrm{B0}}-E_{\mathrm{N0}}\right)^{2}+4\left|\phi\right|^{2}}\end{aligned}
\label{eq1}
\end{equation}

\begin{equation}
\phi=t\left[1+e^{ia\left(-\frac{k_{x}}{2}+\frac{\sqrt{3}k_{y}}{2}\right)}+e^{ia\left(\frac{k_{x}}{2}+\frac{\sqrt{3}k_{y}}{2}\right)}\right]\label{eq2}
\end{equation}

\noindent where $E_{\mathrm{B0}}$ and $E_{\mathrm{N0}}$ are the
onsite energies at the boron and nitrogen atoms, $t$ is the hopping
energy between nearest neighbors, $a$ is the lattice parameter of
hBN, and $\left(k_{x},k_{y}\right)$ is the electron in-plane wavevector.
Fitting of the experimentally measured hBN bands with TBA models is
often avoided in the literature due to the inability to access the
unoccupied conduction band, i.e., to measure the quasiparticle band
gap of epitaxial hBN. However, in order to provide a quantitative
description of the $\pi$ band, we accept a simplistic assumption
that the quasiparticle band gap of monolayer hBN on Ir(111) ($E_{\mathrm{\mathrm{B0}}}-E_{\mathrm{\mathrm{N0}}}$)
corresponds to the optical band gap of hBN monolayer on graphite and
equals 6.1 eV \citep{Elias2019}. After setting the lattice parameter
to $a=2.483$ Å \citep{FarwickZumHagen2016}, fitting the ARPES data
at $\mathrm{\Gamma}$ and K points with Eqs. \ref{eq1} and \ref{eq2}
provides $E_{\mathrm{B0}}=3.73$ eV, $E_{\mathrm{N0}}=-2.37$ eV,
and $t=2.78$ eV. The fitted value of the hopping parameter is in
excellent agreement with the one obtained by fitting the TBA bands
to \textit{ab-initio} calculations of hBN on graphene bilayer \citep{Sawinska2010}.

\subsection*{B. Li coverage calibration}

Alkali metal dispensers used in the experiments exhibit time-dependent
yield \citep{alkali} which depends on the current used for heating
the dispenser, age of the dispenser, and potentially other experimental
factors as well. The amount of alkali atoms released from the dispenser
per unit time increases as the deposition progresses, and this needs
to be taken into account when converting Li deposition time (seconds
or minutes) into Li coverage (in units of $\theta_{\mathrm{max}}$).
In practice, for example, this means that one 5-minute-long deposition
step provides more Li than five consecutive 1-minute-long steps.

We conducted a series of 1- and 2-minute-long deposition steps. By
analyzing the corresponding ARPES spectra and the binding energy of
the $\sigma$ bands, it was found that 13 1-minute-long steps induce
the same $\Delta E_{\mathrm{B}}$ (within the experimental error),
and therefore provide the same amount of Li, as do 4 2-minute-long
steps. By recognizing that Li dispenser yield, $y\left(t\right)$,
is a linear function of time $y\left(t\right)=ct+y_{0}$ for typical
deposition times used in our experiments \citep{alkali}, and by assuming
that the sticking coefficient of Li does not change significantly
during deposition, the equivalence of the two Li coverages requires

\begin{equation}
13\intop_{0}^{1\,\mathrm{min}}\left(ct+y_{0}\right)dt=4\intop_{0}^{2\,\mathrm{min}}\left(ct+y_{0}\right)dt\label{eq3}
\end{equation}

\noindent to be valid. It is then straightforward to show that Li
coverage can be expressed as

\begin{equation}
\theta_{\mathrm{Li}}=\sum_{i}\intop_{0}^{\tau_{i}}y\left(t\right)dt=y_{0}\sum_{i}\tau_{i}\left(\frac{5}{3}\tau_{i}+1\right)\label{eq4}
\end{equation}

\noindent for a sequence of $i$ Li deposition steps of duration $\tau_{i}$.
By defining that a particular combination of deposition steps equals
to $\theta_{\mathrm{Li}}=\theta_{\mathrm{max}}$, all other Li deposition
combinations can be converted to MLs.

\bibliographystyle{elsarticle-num}
\bibliography{refs2}

\end{document}